\documentstyle [aps,draft]{revtex}
\begin{document}
\draft
\title {Self organized criticality in a sandpile model with threshold
dissipation}
\author {Agha Afsar Ali}
\address {Theoretical Physics Group, \\ Tata Institute of Fundamental
          Research, \\ Homi Bhabha Road, Bombay 400 005, India.}
\date {\today}
\maketitle
\begin {abstract} We study a nonconservative sandpile model in one
dimension, in which, if the height at any site exceeds a threshold
value, the site topples by transferring one particle along each bond
connecting it to its neighbours. Its height is then set to one,
irrespective of the initial value.  The model shows nontrivial
critical behavior. We solve this model analytically in one dimension
for all driving rates.  We calculate all the two point correlation
functions in this model, and find that the average local height
decreases as inverse of the distance from the nearest boundary and the
power spectrum of fluctuations of the total mass varies as $1/f$.
\end{abstract}


Self-organized Criticality (SOC) was proposed by Bak, Tang and
Wiesenfeld \cite{btw} to explain the widespread occurrence of fractal
structures and $1/f$ noise in nature. They suggested that extended
dissipative systems evolve to a critical state by a self-organizing
process and proposed the sandpile model as a prototype of SOC
systems. Since then several sandpile-like stochastic cellular automata
models with threshold dynamics have been studied to understand the
origin of criticality. In models with bulk mass conservation
\cite{do,md,ckkp,cc}, the balance of rate of driving and the rate of
loss of particles through the boundary leads to a divergence of the
average avalanche size in the steady state. However, the origin of
criticality in {\it nonconservative} sandpile models is still not well
understood \cite{ff,ofc}. In particular, the nature of nonconservation
which can preserve the criticality of the model are not
known. Analytic proofs of criticality in nonconservative sandpile
models are lacking even in one dimension.

In this paper we study a discrete nonconservative sandpile model
\cite{bk} in which mass dissipation (loss of a particles) occurs if
the height at a site exceeds a threshold value.  In the steady state,
our model organizes such that mass dissipation is minimized. As a
result, it shows nontrivial critical behavior. We calculate its
spatial and temporal correlation functions analytically in one
dimension. The average height profile in the steady state of this
model has a nontrivial power law dependence on the distance from the
nearest boundary. We find that the power spectrum of the fluctuations
of the total mass of the sandpile shows $1/f$ behaviour. These power
laws are robust and do not depend on the details of the model. The
model is critical even for finite driving rate (unlike the forest fire
model\cite{ds}, which is critical only for infinitesimal driving
rate). To our knowledge, this is the first nonconservative model of
SOC which has been shown to be critical for all driving rates.

The model is defined as follows: There is an integer variable $h_i$
(the number of particles), at each site $i$. The particles are added
to the system at randomly selected sites. If $h_i$ exceeds the
threshold height $h_i^c$, then $h_i$ is set to 1 and one particle is
transferred to each neighbor of $i$. This process occurs at all sites
in parallel. We choose $h_i^c$ to be equal to the coordination number
for $i$ inside the bulk and to be greater than the coordination number
for $i$ on the boundary.  Note that this model is not abelian
\cite{do}. In the Abelian sandpile model (ASM), $h_i$ decreases by
$h_i^c$ after the toppling, whereas in this model $h_i$ is set to 1
irrespective of its initial value.

The dynamics of this model resembles the dynamics of the continuous
stick-slip model discussed by Feder and Feder (FF)\cite {ff}.  The FF
model has a real variable $u_i$ at each site $i$, which grows slowly
with time. If $u_i$ exceeds $U_c$, then $u_i$ is set to zero and $u_j$
is increased by 1 for all neighbors $j$ of site $i$. Let us define
$h_i$ as the integer part of $(u_i+1)$ and $h_i^c$ as $(U_c+1)$. Then
clearly, the updating rule can be written in terms of $h_i$ and it is
identical to the toppling rule of our model. However, the external
driving force in FF model is deterministic and not noisy as in our
model.  Note that the argument given above crucially depends on the
fact that the transfer to the neighbours does not depend on the local
variable. Therefore, it does not hold for the earth-quake model
\cite{ofc}, where the transfer is proportional to the local variable.

To illustrate the difference between this model and the ASM, consider
the model on a square lattice and choose an initial configuration in
which the four corners of a given plaquette have all heights equal to
4 (threshold height). Now add a particle at the lower left corner of
the plaquette, say at $ (i,j)$. First $(i,j)$ topples then $(i+1,j)$
and $(i,j+1)$ topple. After this the height at $(i+1,j+1)$ becomes
6. So far there is no dissipation (just as in the ASM). When
$(i+1,j+1)$ topples, unlike in the ASM, 1 particle is lost. Note that
this model is the same as the ASM if the underlying graph does not
have a loop.

In spite of the non-abelian character, the set recurrent
configurations of this model $S_1$ is the same as that of the Abelian
sandpile model $S_2$. To prove this, we first show $S_2 \subset S_1$.
Let us add a particle in $C_0 \in S_2$ and let it relax by the rules
of the ASM. Let the final configuration be $C_1$. Now we add a
particle at the same site in $C_0$ and let it relax by the rules of
this model. If there is no dissipation then the final configuration is
the same as $C_1$. If there is dissipation, then we compensate for the
dissipation by adding particles one by one at sites where dissipation
has occurred and letting the system relax.  This process is repeated
until no site is left where dissipation has occurred. Since in the ASM
the final configuration does not depend on the order of the topplings,
this sequence of additions and relaxations in this model must lead to
$C_1$. Thus, in this model, there is a finite probability of
transitions from one recurrent configuration of the ASM to the
another. This implies that $S_2 \subset S_1$. One can easily check
using the argument of Ref. \cite{do} that the forbidden configurations
of the ASM are also forbidden in this model. Therefore, $S_{1} =
S_{2}$. However, unlike the abelian case, the probabilities of
occurrence of different recurrent configurations in the steady state
of this model need not be equal.

Let us analyse this model in one dimension. The model on a simple
linear chain is not different from the ASM, because this lattice has
no loops. Hence, we consider this model on decorated one dimensional
chains, formed by joining unit cells which have loops (see Fig. 1).
We have earlier studied ASM's on this type of chain and found
interesting finite-size scaling behavior \cite{ad}, which differs from
that seen in simple linear chains\cite{rs}.

Consider first the chain of doublets (case A, Fig. 1). We label the
doublets by integers $i = -l $ to $l$ and the sites inside the doublet
by $j = 1$ to 2. The size of the chain $L=2l+1$. We generalize the
rule such that after the toppling one particle is transferred along
each bond connecting the site to its neighbors, and take the threshold
height equal to 3 for all sites. The recurrent configurations of this
model are characterized using the burning algorithm as in the ASM This
algorithm is specified by the following rule: A site is burnt if its
height is greater than the number of bonds joining it to its unburnt
neighbours (see
\cite{do}, for detail). A stable configuration is recurrent if and
only if all the sites are eventually burnt.

In this algorithm the sites can be burnt in any order. We let the
burning start from the left boundary and hold right boundary unburnt.
The point at which burning from this boundary stops is called the
break point (BP). Afterwards, the right boundary is burnt and
subsequently the remaining sites. Thus the allowed values of
$(h_{i1},h_{i2})$ for $i$ to the left of the BP are $(3,3)$ and
$(3,2)$, and those for $i$ to the right of the BP are $(3,3)$ and
$(2,3)$. The allowed values at the BP are $(1,3)$, $(3,1)$ and
$(2,3)$.

In any side the avalanche spreads to the first doublet which cannot be
burnt from that side. For example, if the particle is added to the
left of the BP the avalanche spreads to the BP on the right and to the
first doublet of type $(3,2)$ on the left. For a typical avalanche the
distance between the site at which the particle is added and the BP is
$O(L)$. Therefore, the spread of avalanche is $O(L)$.

When the avalanche crosses a doublet of type $(3,3)$, one particle is
dissipated. Thus, each avalanche wipes out $(3,3)$ type of doublets
from a region of order $L$, leaving a small density of the $(3,3)$
type doublets. This minimizes the possibility of dissipation. Also,
since the steady state configurations are dominated by $(3,2)$ type
doublets ($(2,3)$ type doublets) on the left (right) of the BP, the
leftward (rightward) spread of the avalanche starting from the left
(right) of the BP is small (see Fig. 2).

The BP shows an interesting stochastic dynamics. After the avalanche,
it moves towards the starting point of the avalanche, by a distance of
order 1. If $i$ is the position of the BP then the probability that an
avalanche starts from the left of the BP is $(l+i)/(2L)$ to leading
order (ignoring the density of $(3,3)$ type of doublets), and that it
starts from the right of the BP is $(l-i)/(2L)$. Hence, the mean
displacement of the BP after the avalanche scales as $i/(2L)$. The
fluctuation about this mean value is of order 1.  The dynamics of the
BP can thus be described by an equation of the following form
\begin{equation} {dx \over dt} = - {x \over 2L} + \eta (t)
\label{eq1}
\end{equation} where $x$ is the scaled variable $i/L$, and $dt$ is of
the order of the time interval at which avalanches hit the break
point. The noise $\eta (t)\sim 1/L$, and is $\delta$-correlated. From
the above equation it follows that the number of avalanches required
to reach the steady state is of order $L$, and the asymptotic
distribution of the BP goes as $\exp (-cx^2L)$, where $c$ is a
constant. The width of this distribution can be ignored in the large
$L$ limit. A typical avalanche then starts from the point at which
particle is added (source point) and terminates at the center. If the
source point is $i$ then the linear extension $s$ and the duration $t$
of an avalanche are equal to $|i-L/2|$. Averaging over $i$ we get
\begin{equation}
\hbox{Prob}\,(X) \sim 2/L~~\hbox{for}~~X<L/2~,
\label{eq3}
\end{equation} and 0 otherwise, where $X=t,s$.

Now we calculate the correlation functions in this model. The model
has two intrinsic time scales, the time taken for one toppling
(defined as one time step), and the time interval $T$ between two
consecutive addition of particles.  Let us first consider the case,
$T>L$, such that there is no overlap between two consecutive
avalanches (see Eq.  (\ref{eq3})).

It is convenient to use a two state variable $s_i$, because there are
only two allowed configurations of each doublet (except the break
point). For the $(3,3)$ type doublet, $s_i = 1$, otherwise $s_i =
0$. Consider the evolution $s_i$ for $\sqrt{L} \ll i
\ll l$. We can assume that the BP is on left of $i$. If the avalanche
starts from the right of $i$, then it crosses $i$ to reach the BP,
setting $s_i=0$. The probability of this transition goes as
$(l-i)/(2L)$.  The transition of $s_i$ from 0 to 1 occurs, if the
particle is added at the left site of the $i$-th doublet, or if it is
added to the left of the $i$-th doublet, but the avalanche reaches the
$(i-1)$-th doublet so that the left site of $i$-th doublet receives a
particle.  The probability of this transition goes as $1/L$.  The
corresponding probabilities for $-l
\ll i \ll -\sqrt{L}$, can be obtained by replacing $i$ by $(-i)$ and
$l$ by $(-l)$. Thus the probability $P_n(i)$, that $s_i=1$ after
additions of $n$ particles, satisfies the following equation (to
leading order)
\begin{equation} P_{n+1}(i) = {1 \over L} \left[1-P_n(i)\right] -
\left( 1-{r(i)\over 2L}\right) P_n(i)~~,
\label{eq4}
\end{equation} where $r(i)$ is the distance of $i$ from the nearest
boundary. A straightforward calculation using the above equation gives
\begin{equation}
\langle s_i \rangle \sim 2/r(i)~~,\\
\label{eq5}
\end{equation} and the autocorrelation
\begin{equation}
\langle s_i(t_0)s_i(t_0+t)\rangle \sim {1\over
r(i)}\left(1-\tau_i^{-1} \right)^{t / T}~~,
\label{eq51}
\end{equation} for $t \gg T$, where $\tau_i ={2L \over r(i)}$ and
$\langle \dots \rangle$ denotes average over $t_0$. The value of $s_i$
at time $t$ is denoted by $s_i(t)$. Note that the correction to
Eq.(\ref{eq4}) coming from the nonzero density of $s_i$
(Eq. (\ref{eq5})) vanishes in the large $L$ limit.  The steady state
of this model is not translationally invariant, but is self similar in
space, i.e. the density and the relaxation time $\tau_i$ depend on its
distance from the nearest boundary as a power law. In Fig. 3, we plot
the density of $s_i$ against the distance from the left boundary. Note
the excellent agreement of our numerical data with Eq. (\ref{eq5}).

To calculate the correlation between $s_i(t)$ and $s_j(t^\prime)$,
consider the evolution of the joint probability of $s_i$ and
$s_j$. There are four possible values of $(s_i,s_j)$. Let the
four-vector, $P_n(i,j)$, denote the joint probability of $(s_i,s_j)$
after addition of $n$ particles. Its evolution can be described by the
following linear equation (to leading order)
\begin{equation} P_{n+1}(i,j) = G(i,j) P_n(i,j)~~,
\label{eq6}
\end{equation} where $G(i,j)$ is the $(4 \times 4)$ matrix giving the
transition probabilities of $s_i$ and $s_j$.  This can be calculated
in exactly the same way as the transition probability of $s_i$
calculated above. For example, (1,1) goes to (0,0) if the avalanche
passes over both the doublets $i$ and $j$. If $\sqrt{L} \ll i \ll j
\ll l$ then the probability of this transition goes as
$r(j)/(2L)$. Other transition probabilities can also be calculated in
the same way. From Eq. (\ref{eq6}) one can find the steady state
correlation functions. For $i$ and $j$ on the same side of the center
of the chain
\begin{equation}
\langle s_is_j \rangle _c ~\sim {1\over r(i)^2}~,
\label{eq7}
\end{equation}
\begin{equation}
\begin{array}{l}
\langle s_i(t_0)\,s_j(t_0+t) \rangle _c \sim {1\over
r(i)^2}\left(1-\tau_j^{-1}
\right)^{t / T}~,\\
{}~~~~~~~~~~~~~~~~~~~~~~~~~~~~~~~~~~~~\hbox{for}~r(i)>r(j)~,
\label{eq8}
\end{array}
\end{equation}
\begin{equation}
\begin{array}{l}
\langle s_i(t_0)\,s_j(t_0+t) \rangle _c \sim -{1\over
r(j)(r(j)-r(i))}\left(1-\tau_i^{-1} \right)^{t / T}~,\\
{}~~~~~~~~~~~~~~~~~~~~~~~~~~~~~~~~~~~~~~~\hbox{for}~r(i)<r(j).
\end{array}
\label{eq81}
\end{equation} The subscript $c$ refers to the connected part of the
correlation function. If the $i$ and $j$ are on different sides of the
center then the correlation can be ignored. Note that, in Eq.
(\ref{eq8}), the first factor gives the equal time correlation, and
the second factor gives the relaxation of $s_j$. In Eq. (\ref{eq81})
the $s_i(t_0)$ and $s_j(t_0+t)$ are anti-correlated because an
avalanche which makes $s_i=1$ sweeps over site $j$, setting $s_j=0$.

The autocorrelation of mass (sum of heights) can be obtained using
Eq. (\ref{eq51}), (\ref{eq8}) and (\ref{eq81}). The power spectrum of
mass fluctuation $S(f)$, is obtained by taking the real part of the
Fourier transform of the autocorrelation
\begin{equation} S(f) \sim {1 \over f}~~,~~\hbox{for}~~{2\pi \over TL}
\ll f \ll 2\pi~~.
\label{eq9}
\end{equation} In Fig. 4, we show the log-log plot of the numerical
value of $S(f)$ versus $f$. This is obtained by taking the square of
the Fourier transform of the time sequence of length 5000, for a chain
of size 5000. We find that at low frequencies the data shows $1/f$
behaviour. The scaling region increases with the increase in the
lattice size. Since we have taken only the slowest relaxation mode in
Eq. (\ref{eq8}), (\ref{eq81}) and (\ref{eq51}), the agreement is not
good in the high frequency regime.

Consider now the case in which particles are added at a faster rate,
i.e., $1 \ll T \ll L$. We note that the interaction between the
avalanches can be ignored because the avalanches propagate with the
same velocity towards the BP (see Fig. 2). Thus the steady state
properties of the model as described by Eq. (\ref{eq7}) and
(\ref{eq5}) and the autocorrelation given in Eq. (\ref{eq51}) remain
unchanged in this case. Since the avalanche takes $2|j-i|$ time steps
to travel from $i$ to $j$, there is no anti-correlation between
$s_i(t_0)$ and $s_j(t_0+t)$ for $t < 2|j-i|$ (see
Eq. (\ref{eq81})). However, the contribution to the mass
autocorrelation coming from the this term can be ignored for $|i-j| >
\sqrt{L}$, or for $t > \sqrt{L}$. Thus for $f < 2\pi T/\sqrt{L}$, the
power spectrum $S(f) \sim 1 / f$.

To check the robustness of our results we have studied the model on
the diamond chain (case B in Fig. 1). An avalanche in this case has
same structure as in case A, i.e., it spreads to the BP on one side,
to a distance of $O(1)$ on other side and the BP is confined in a
region of $O(\sqrt{L})$. As a result, the power laws of the average
height and the power spectrum of mass fluctuation are also the same.

The two special doublet configurations in this model, which stop the
avalanches, are reminiscent of the `trough' and `trap' of the
`singular diffusion' type model studied by Carlson et al \cite{cc}. In
that case too there is a power law decay of the height profile which
follows from the singularity of the effective diffusion coefficient of
the particles as a function of the coarse grained density
\cite{ckkp,k}.  However, in our model the mechanism of
self-organization is different as the diffusion coefficient does not
diverge.

To summarize, we have determined exactly the critical behavior of a
non-abelian nonconservative sandpile model in the one dimension. The
critical steady state shows spatial structures. We determine time
dependent correlation functions for finite driving rate, and show that
fluctuations of the mass of the sandpile have a $1/f$ spectrum.

I thank Prof. Deepak Dhar for his useful suggestions and in
particular, for pointing out that the set of the recurrent
configurations of this models is same as that of the ASM. I thank
Gautam I.  Menon and P. Lakdawala for a critical reading of the
manuscript.

{\bf Captions}

Fig. 1: The one dimensional chains formed by joining (A) doublets, (B)
diamonds and (C) single sites.

Fig. 2: The evolution the sandpile. The dot denotes an toppling event.

Fig. 3: The log-log plot of the mean height of a doublet (minimum
height is subtracted) versus the distance from the nearest boundary
$r$. The solid line shows the analytic expression.

Fig. 4: The log-log plot of power spectrum of total mass fluctuation
$S(f)$ vs frequency $f$. For comparison the theoretical curve is shown
as the solid line.

\end {document}
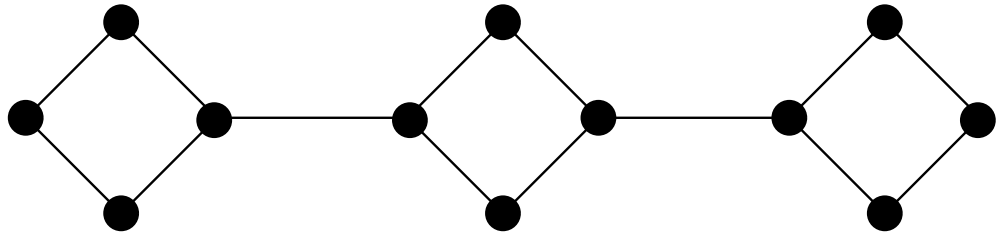
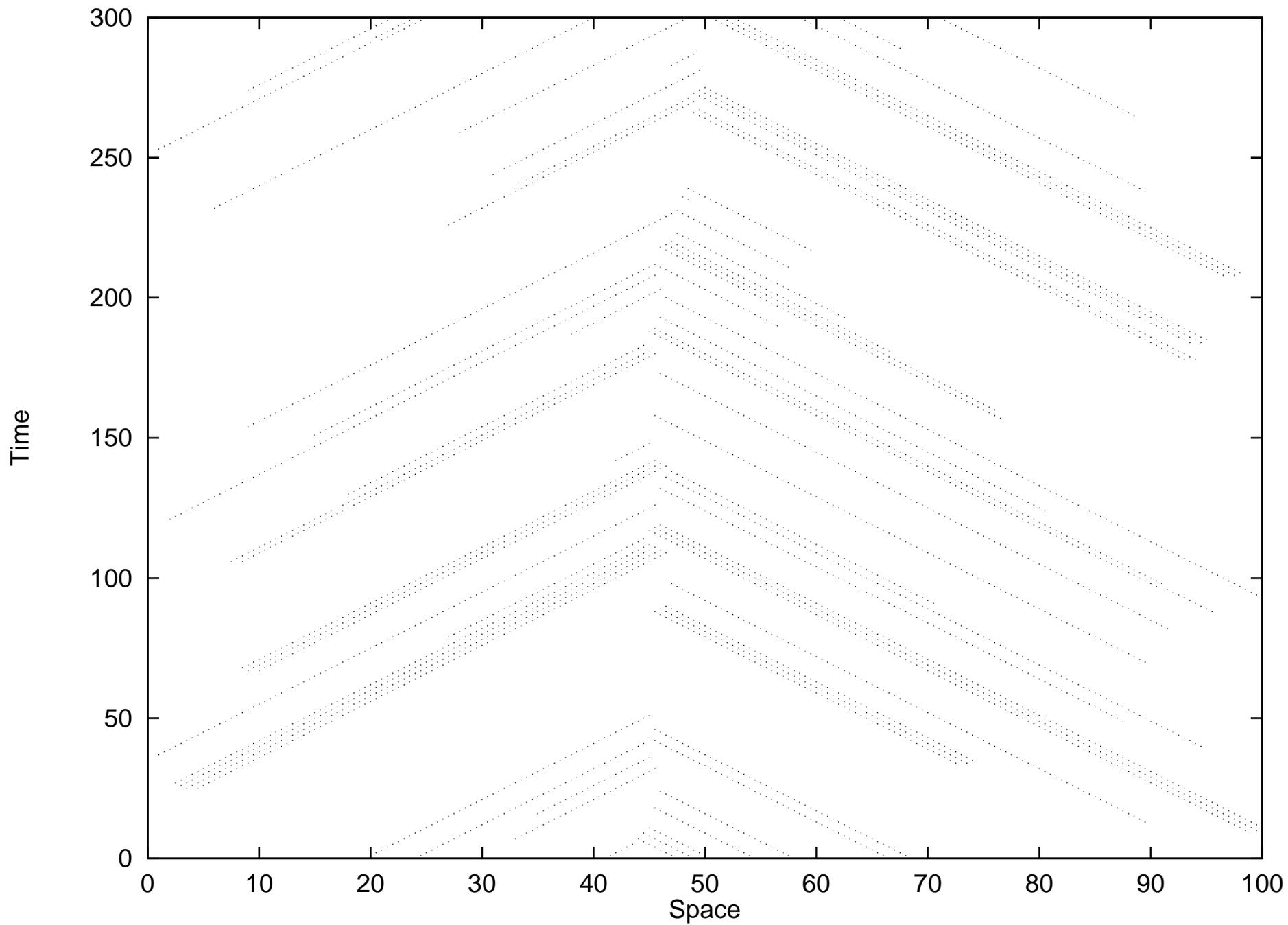
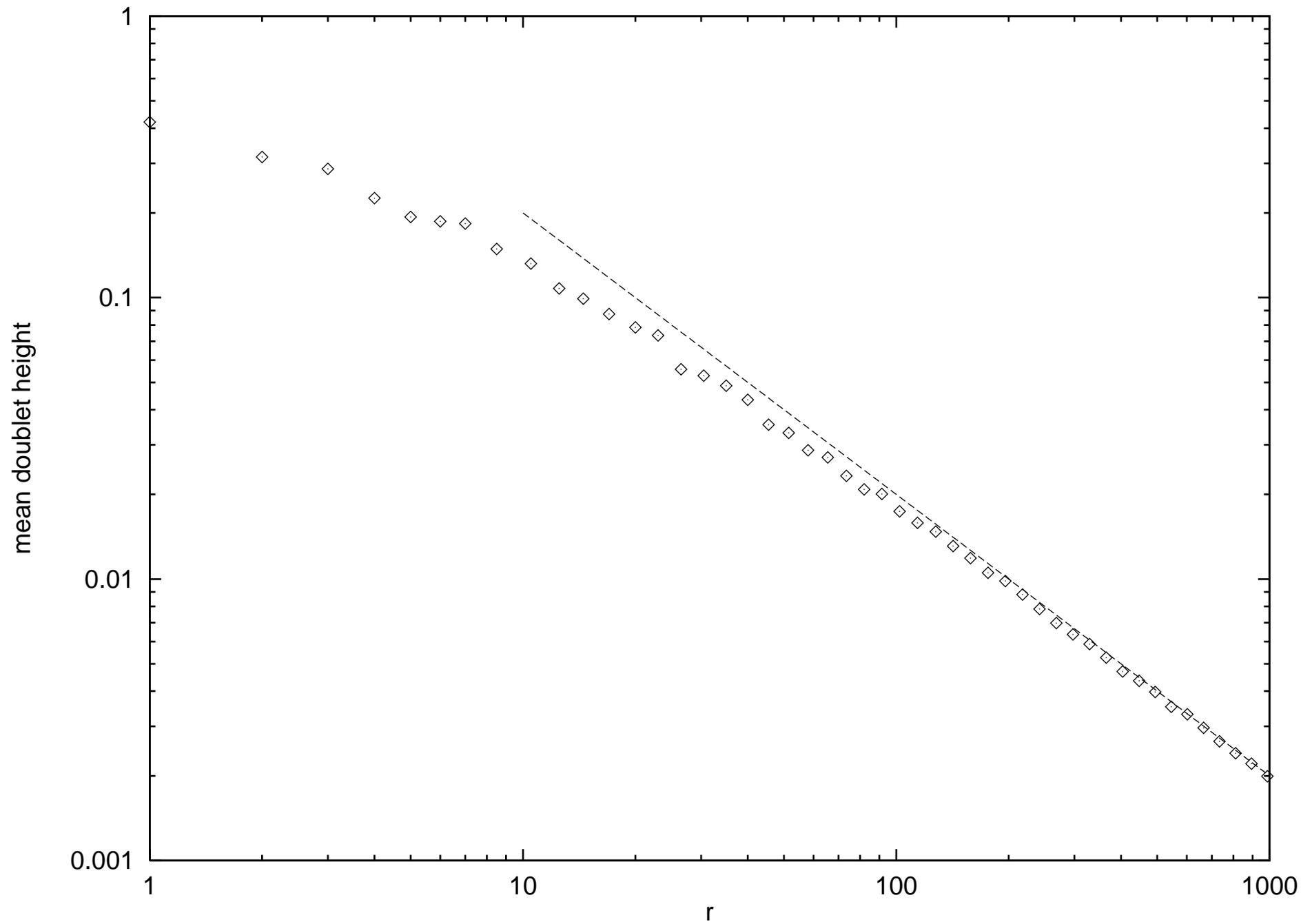
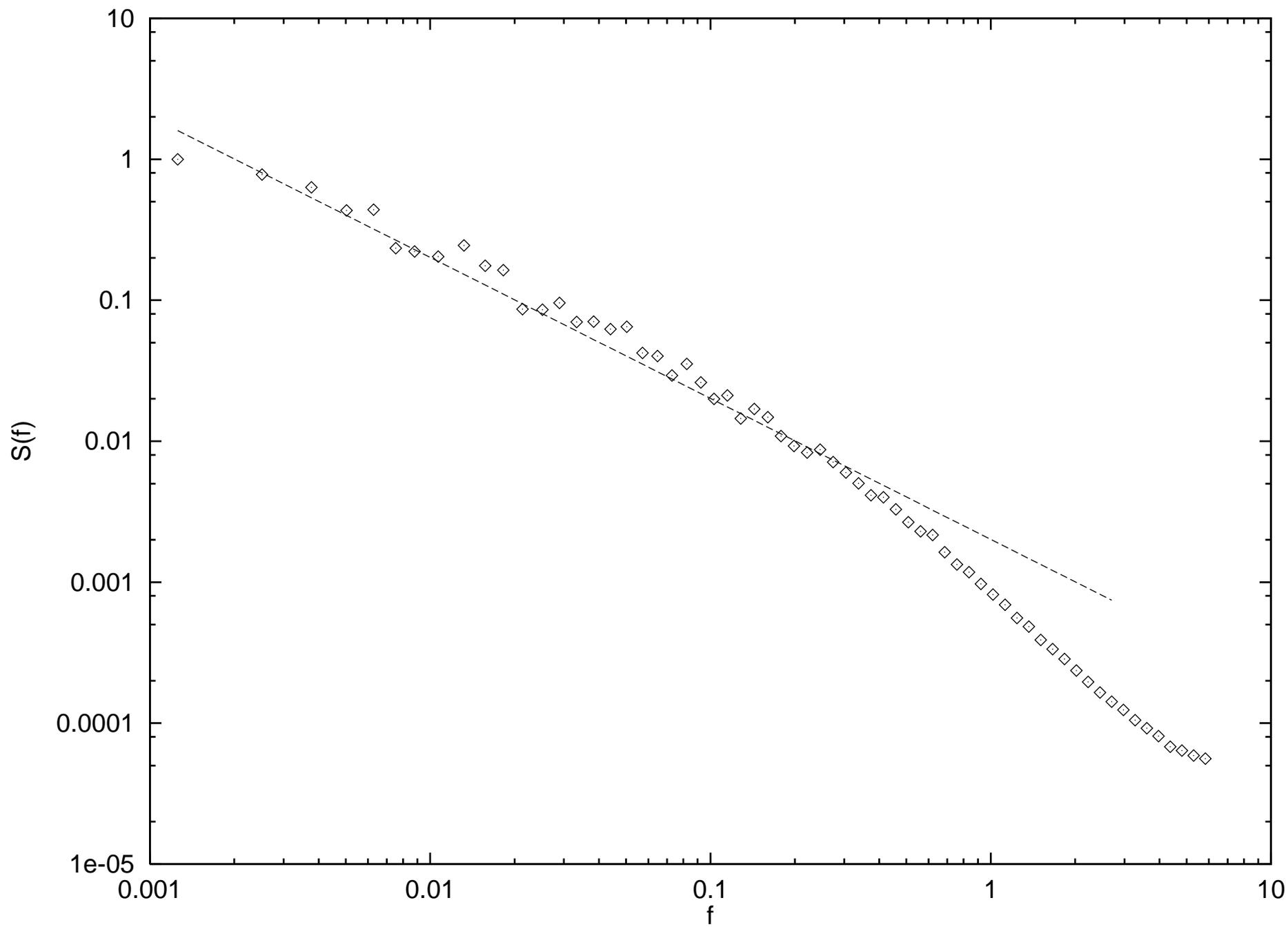
